\documentclass[]{mnras}

\title[{\sl Kepler} observations of KIC 9202990]{Continuous `stunted'
  outbursts detected from the Cataclysmic Variable KIC 9202990 using {\kep} data}

\author[]
{Gavin Ramsay$^{1}$, Pasi Hakala$^{2}$, Matt A. Wood$^{3}$, Steve B. Howell$^{4}$
Alan Smale$^{5}$,\and Martin Still$^{4,6}$, Thomas Barclay$^{4,6}$\\
$^{1}$Armagh Observatory, College Hill, Armagh, BT61 9DG, UK\\
$^{2}$Finnish Centre for Astronomy with ESO (FINCA), University of Turku,
V\"{a}is\"{a}l\"{a}ntie 20, FI-21500 PIIKKI\"{O}, Finland\\
$^{3}$Department of Physics and Astronomy, Texas A\&M University-Commerce, 
Commerce, TX 75429-3011, USA\\
$^{4}$NASA Ames Research Center, Moffett Field, CA 94095, USA\\
$^{5}$NASA/Goddard Space Flight Center, Greenbelt, MD 20771, USA\\
$^{6}$Bay Area Environmental Research Institute, Inc., 560
Third St. West, Sonoma, CA 95476, USA\\
}

\date{Accepted 2015 October 26.  Received 2015 October 24; in original form 2015 August 1}

\begin{document}
\outer\def\gtae {$\buildrel {\lower3pt\hbox{$>$}} \over 
{\lower2pt\hbox{$\sim$}} $}
\outer\def\ltae {$\buildrel {\lower3pt\hbox{$<$}} \over 
{\lower2pt\hbox{$\sim$}} $}
\newcommand{\Msun} {$M_{\odot}$}
\newcommand{\Rsun} {$R_{\odot}$}
\newcommand{\kep}{\it Kepler}
\newcommand{\swift}{\it Swift}
\newcommand{\Porb}{P_{\rm orb}}
\newcommand{\nuorb}{\nu_{\rm orb}}
\newcommand{\eplus}{\epsilon_+}
\newcommand{\eminus}{\epsilon_-}
\newcommand{\cd}{{\rm\ c\ d^{-1}}}
\newcommand{\MdotL}{\dot M_{\rm L1}}
\newcommand{\Ldisk}{L_{\rm disk}}
\newcommand{\src}{KIC 9202990}
\newcommand{\ergscm} {ergs s$^{-1}$ cm$^{-2}$}

\maketitle
\begin{abstract}
Based on early {\kep} data, {\O}stensen et al. (2010) found that
{\src} showed a 4 hr and a two-week photometric period. They suggested
the 4 hr period was a signature of an orbital period; the longer
period was possibly due to precession of an accretion disk and {\src}
was a cataclysmic variable with an accretion disk which is always
  in a bright state (a nova-like system). Using the full {\kep}
dataset on {\src} which covers 1421 d (Quarter 2--17), and includes 1
min cadence data from the whole of Quarters 5 and 16, we find that the
4 hr period is stable and therefore a signature of the binary orbital
period. In contrast, the 10--12 d period is not stable and shows an
amplitude between 20--50 percent. This longer period modulation is
similar to those nova-like systems which show `stunted'
outbursts. We discuss the problems that a precessing disk model
  has in explaining the observed characteristics and indicate why we
  favour a stunted outburst model.  Although such stunted events are
  considered to be related to the standard disk instability mechanism,
  their origin is not well understood.  {\src} shows the lowest
amplitude and shortest period of continuous stunted outburst systems,
making it an ideal target to better understand stunted outbursts
  and accretion instabilities in general.
\end{abstract}

\begin{keywords}
Stars: individual: {\src}: Stars: binaries -- Physical data and
processes: accretion and accretion discs -- instabilites

\end{keywords}

\section{Introduction}

The {\kep} satellite observed the same 115 square degree field of
view, just north of the Galactic plane in Cygnus and Lyra, between
April 2009 and May 2013. It allowed virtually uninterrupted photometry
of more than 150,000 stars with 30 min cadence and 512 stars with 1
min cadence at any one time.  Although the prime goal of the {\kep}
mission was to discover Earth-sized planets orbiting the host stars
habitable zone (e.g. Borucki, et al. 2013), it has led to a revolution
in the field of asterseismology across the HR diagram (e.g. Chaplin et
al. 2014).

{\kep} has also provided a unique opportunity to study accreting
sources such as cataclysmic variables (CVs) which show flux
variations over timescales ranging from seconds to years or even
decades. Early observations such as those of V344 Lyr (Still et
  al. 2010) showed the potential of {\kep} to address key questions
  regarding the nature of the outbursts seen in CVs. CVs contain a
white dwarf which is accreting material through Roche Lobe overflow
from a late-type star. The observed characteristics of the system is
largely set by the binary orbital period and the strength of the
magnetic field of the white dwarf. {\kep} has observed dozens of CVs,
some of which were known prior to its launch (see Howell, et
al. 2013), while others were discovered purely by chance (e.g. Barclay
et al. 2012, Brown et al. 2015).

Very early in the mission, {\O}stensen et al. (2010) presented
  initial results of {\sl Kepler} observations of a sample of known or
  suspected compact pulsators.  One of these sources, {\src}, showed
a strong modulation on a timescale of two weeks together with a second
much shorter period ($\sim$4 hrs) superimposed. They presented an
optical spectrum which showed Balmer lines in absorption but filled in
with emission and appeared to be disc-dominated and similar to
nova-like CVs. {\O}stensen et al. (2010) suggested the 4 hr
  period represented the binary orbital period which would be typical
  of nova-like CVs which lie predominately above the 2--3 h period gap
  (see G\"{a}nsicke et al 2009 for an overview of the orbital period
  distribution of CVs) and has a high mass transfer rate and an
  accretion disc in a bright, steady state (see Dhillon 1996 for a
  review of nova-like CVs and also Aungwerojwit et al. 2005). They
suggested that the 4 hr period was the orbital period and the longer
period was due to a precessing disc. McNamara, Jackiewicz \& McKeever
(2012) noted it was a possible pulsating B star based on an analysis
of the light curve.

\begin{figure}
\begin{center}
\setlength{\unitlength}{1cm}
\begin{picture}(6,5.5)
\put(10,-0.7){\includegraphics{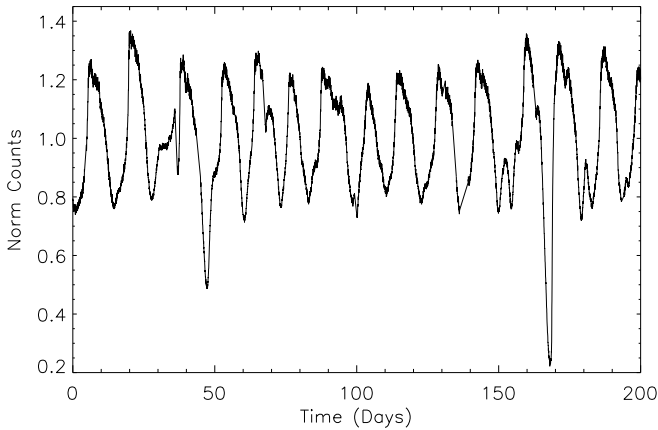}}
\end{picture}
\end{center}
\caption{As an example of the flux variations of {\src} on a timescale
  of tens of days, we show a light curve covering 200 d (MJD
  55872--56072, 2011 Nov 07 -- 2012 May 25). Note the prescence of two
  prominent dips at 45 d and 168 d from the start.}
\label{lclight} 
\end{figure}

With the {\kep} mission now having ended as it was first envisaged, we
have taken all the Kepler data on {\src} and set out to determine if
the nova-like CV designation can be supported by the much more
extensive dataset than was available to {\O}stensen et al. (2010).

\section{{\src}}

{\src} has a position $\alpha=18^{\rm h}56^{\rm m}08.1^{\rm s}$
$\delta = +45\degr 37^{'}40.1^{''}$ J2000.0 (taken from the Kepler
Input Catalog). It was not detected in the {\sl ROSAT} All-Sky Survey,
but does not appear to have been in the field of any pointed {\sl
  XMM-Newton} or {\sl Chandra} observations. It was observed in
  the {\sl Kepler INT Survey} (KIS, Greiss et al. 2012a,b) which
  obtained $Ugri$H$_{\alpha}$ photometry of the majority of sources in
  the {\sl Kepler} field. The mean photometry and colours of {\src}
  are $g$=15.39$\pm$0.04, $U-g$=--0.62, $g-r$=0.12, $r-i$=0.15 and
  $r-$H$_{\alpha}$=0.23. The latter colour index is consistent with
  H$\alpha$ in absorption. The UBV Survey of the {\kep} field
(Everett, Howell \& Kinemuchi 2012) also indicate a very blue source:
$B-V$=0.09, $U-B=-$0.70. These colours are consistent with other
CVs. {\src} has also been detected using the Catalina Real-time
Transient Survey (Drake et al. 2009). There are 69 photometric data
points spread over 8 years showing a mean of $V\sim15.0\pm0.2$ and a
range of $V\sim$14.6--15.5.

\section{Kepler Photometric Observations}

\begin{figure}
\begin{center}
\setlength{\unitlength}{1cm}
\begin{picture}(12,5)
\put(10.,-0.7){\includegraphics{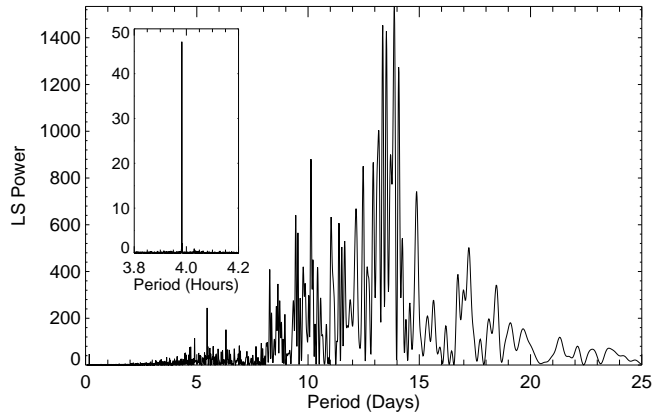}}
\end{picture}
\end{center}
\caption{The Lomb Scargle power spectrum derived using the full LC
  light curve of {\src}. The many peaks are indicative of the
    complex nature of the long period (the window function does not
    show significant side-lobes). In the smaller panel we show the
    power spectrum centered near the period of the orbital period
    (3.98 hrs).}
\label{lcpower} 
\end{figure}

The vast majority of targets in the {\kep} field were observed in {\it
  long cadence} (LC) mode, where the effective exposure is 27.1 min. A
very small number of targets (the targets could be changed every
month) were observed in {\it short cadence} (SC) mode, where the
effective exposure is 54.2 s. As the satellite was rotated every 3
months to ensure the solar array was effectively pointed to the Sun,
there are short data gaps when data was downloaded from the
satellite. {\src} was observed for 16 `Quarters' (Q2--17), giving
almost 4 years of near continuous data between June 2009 and May 2013,
and it was observed for 7 months using SC mode (Q2/3, Q5/1--3 and
Q16/1--3). After the raw data are corrected for bias, shutterless
readout smear, and sky background (see Jenkins et al. 2010),
time series are extracted using simple aperture photometry (SAP).

The SAP data were extracted from the data files downloaded from
  the MAST archive\footnote{http://archive.stsci.edu/kepler} and 
filtered to remove data from time intervals when the data
may have been compromised, for instance by enhanced Solar activity,
(we filtered data points so that `SAP\_QUALITY=0'). We then normalised
each quarter of data so that the mean count rate was unity. To remove
systematic trends in the data we used the task {\tt kepcotrend} which
is part of the {\tt PyKE} software (Still \& Barclay
2012){\footnote{http://keplergo.arc.nasa.gov/PyKE.shtml}}. We then
applied a small offset so that there are no discrete jumps in flux
between the different quarters of data. The same steps were applied to
the SC data.

\subsection{Long Cadence Data}

To highlight the photometric variability of {\src} on a timescale of
tens of days, we show 200 d of LC data of {\src} in Figure
\ref{lclight}. As first reported by {\O}stensen et al. (2010) there is
a prominent modulation on a timescale between 10 and 12 d and a
full-amplitude which varies between 20 and 50 percent. We also note
the presence of two clear `dips' in the light curve. We show the Lomb
Scargle power spectrum of the full LC light curve in Figure
\ref{lcpower}, which covers 1421 d, but has short gaps every 3
months. The power spectrum is complex but shows four prominent peaks
corresponding to periods between 13--14 days. This implies the large
modulation seen in Figure \ref{lclight} is not strictly periodic but
quasi-periodic (we call this the `long' period).

We also determined the length of each cycle by simply determining the
time of maximum flux of each long period cycle. The distribution of
the duration of each cycle can be approximated with a Gaussian
function with a mean duration of 12.5 d with a FWHM of $\sim$4 d.

We also searched for periodic behaviour on timescales greater than the
long period. Since long term trends maybe removed when we normalise
the light curve on a quarter basis, we used the light curve which was
de-trended but not normalised. There is clear peak in the power
spectrum at $\sim$185 d which is due to a modulation with a
full-amplitude of $\sim$10 percent in the light curve. (This period is
not seen in the power spectrum of the normalised light
curve). However, we note that this period is half the orbital period
of the {\kep} satellite. B\'{a}nyai et al. (2013) and Hartig et
al. (2014) present studies of {\kep} observations of long period
variables and find some evidence that year long periods maybe
artifacts in the SAP derived light curve. We therefore add some
caution regarding whether the 185 d modulation is intrinsic to {\src}.

\subsection{Short Cadence Data}

\begin{figure}
\begin{center}
\setlength{\unitlength}{1cm}
\begin{picture}(6,6.5)
\put(9,-0.5){\includegraphics{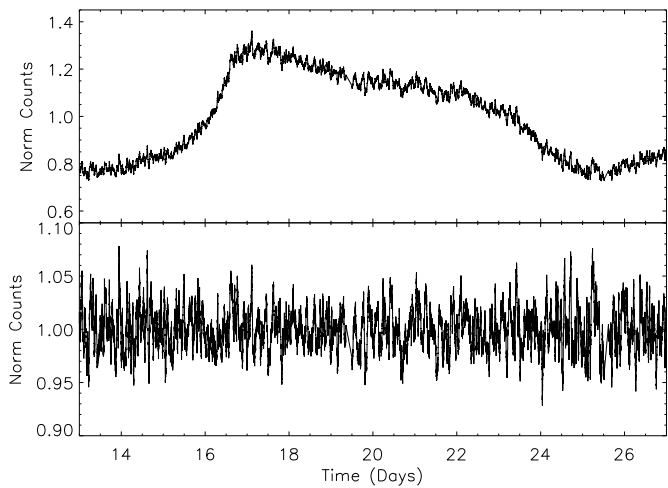}}
\end{picture}
\end{center}
\caption{The SC light curve of {\src}, derived from Q5 data, shows one
  long cycle (top panel). The lower panel shows the light curve after
  the effects of this long cycle have been removed.}
\label{sclight} 
\end{figure}

As noted by {\O}stensen et al. (2010), in addition to the long
  period, high amplitude modulation, there is also a short period, low
  amplitude modulation present in the light curve of {\src}. Although
  the cadence of the LC data is much lower than the SC data, the
  signature of the shorter period is seen in the power spectrum of the
  LC dataset (Figure \ref{lcpower}). However, the SC gives a much
  higher resolution of the short period modulation. This can be seen
in Figure \ref{sclight} where we show one complete cycle of the long
period and the resulting light curve after this modulation has been
removed using the {\tt PyKE} task {\tt kepdetrend}. We derived the
Lomb Scargle power spectrum of each month of SC data: each power
spectrum shows a dominant period at 3.98 hrs. We then folded each
month of data (and also combined data from each quarter) on the
ephemeris:

\begin{equation}
T_{o} = BMJD 55063.816924 + 0.1659404 E \noindent
\end{equation}

\begin{figure}
\begin{center}
\setlength{\unitlength}{1cm}
\begin{picture}(6,11)
\put(-1,-0.8){\includegraphics{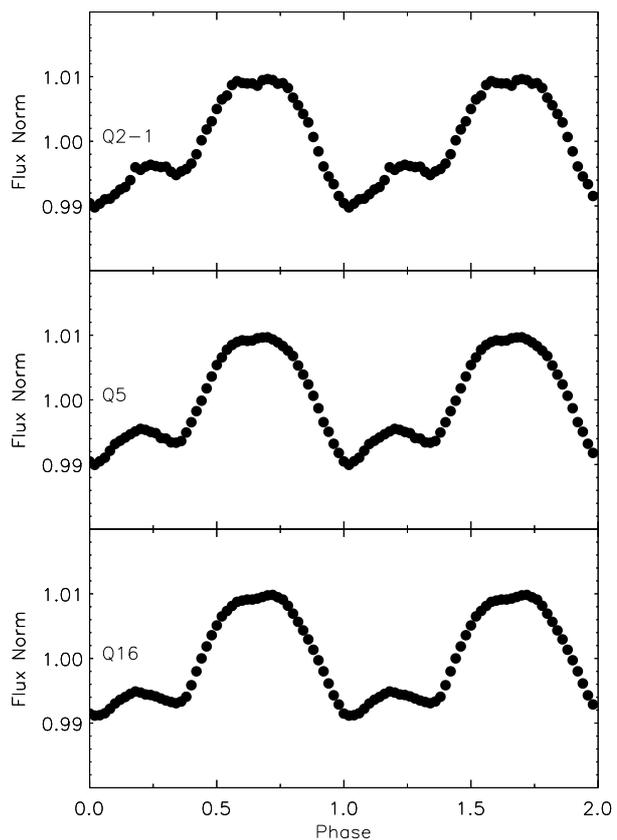}}
\end{picture}
\end{center}
\caption{The SC data of {\src} taken from three separate quarters in
  time have been folded on the ephemeris shown in eqn 1. The
  repeatability of the resulting profiles indicates that the 4 h
  period is a signature of the orbital period.}
\label{scfold} 
\end{figure}

As can be seen from the folded light curves (Figure \ref{scfold}) the
shape and phase of the light curves are virtually identical. We
  fit the LC light curve using a sinusodial wave with a period of
  0.1659404 d and obtained a set of residuals using a 10 cycle
  fit. Although the residuals show a two week quasi-period which is
  almost certainly the result of the changing shape of the orbital
  period profile (see \S \ref{long}), there is no systematic change in
  the O-C residuals over the course of the LC light curve. This
strongly suggests that the 0.1659404 d period is extremely stable and
must be the signature of the binary orbital period (as suggested by
{\O}stensen et al. 2010).

\section{Orbital and long period variations}
\label{long}

\begin{figure*}
\begin{center}
\setlength{\unitlength}{1cm}
\begin{picture}(10,9)
\put(-3,-1.2){\includegraphics{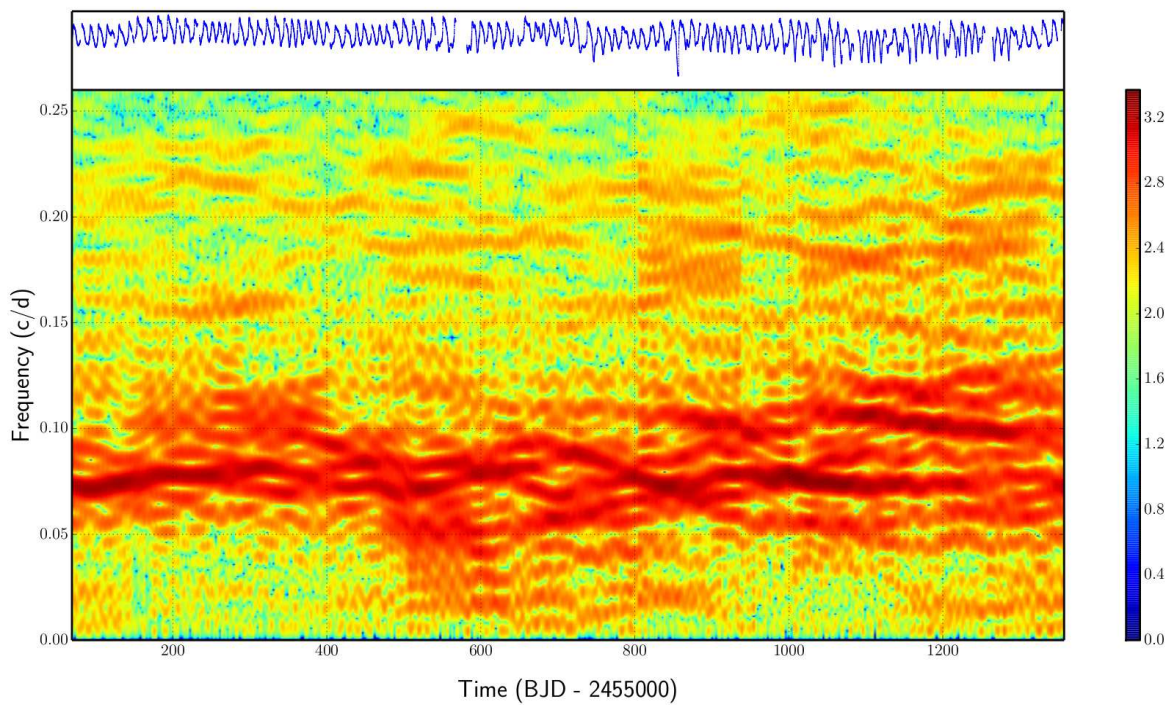}}
\end{picture}
\end{center}
\caption{Using a sliding Lomb Scargle Periodogram we show how the long
  period varies over the course of the observations. The top panel
  shows the {\kep} light curve of {\src} and the right hand panel
  shows the power of the periodogram. The frequency of the dominant
  peak in the periodogram clearly changes over time.}
\label{spectrogram} 
\end{figure*}

The forest of peaks in the periodogram of {\src} (Figure
\ref{lcpower}) suggests that the long period is not stable.  In order
to investigate this further, we have performed more detailed
investigations. The simplest test is to obtain a Lomb Scargle
periodogram of relatively short sections of data in a sliding
manner. We show the resulting 'spectrogram' in Figure
\ref{spectrogram}: it confirms that the long period is not stable
and shows other periods that appear to be transient in nature.

\begin{figure}
\begin{center}
\setlength{\unitlength}{1cm}
\begin{picture}(6,6)
\put(8.,-1.2){\includegraphics{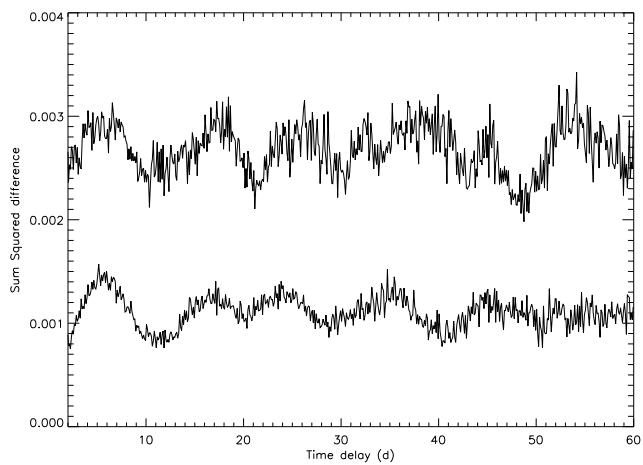}}
\end{picture}
\end{center}
\caption{We show the results of measuring the difference in the
  orbital light curves as a function of time for Q5 SC data (top) and
  Q16 SC data (bottom). This indicates the longer period is modulated
  on a quasi-period of 10--14 d (see text for details).}
\label{q5and6} 
\end{figure}

We then carried out a more sophisticated analysis using both the Q5
and Q16 SC datasets which are both approximately 90 d in length. We
perform the same analysis on each light curve separately, which aims
to measure the similarity of the orbital light curve shape as a
function of their separation in time. Our analysis proceeds as
follows: i) Pick two non-overlapping pieces of light curves, each 2 d
(giving 12 orbital cycles) long, separated by a random amount of time
(from 2 d up to the length of the dataset); ii) Fold these two
subsections of light curves over the 4 h orbital period and bin each
of these into 50 orbital phase bins; iii) Compute the sum of squared
differences of the binned light curves and iv) repeat this 100,000
times. As a result, we have a measure of the difference between
orbital light curve shapes as a function of their separation in
time. Finally, we have binned the results in 0.1 d time bins: in both
datasets we see a minimum near 10--12 d separation and its multiples
(Figure \ref{q5and6}). This strongly suggests that the orbital light
curve shape is indeed modulated on a 10--12 d period, which
corresponds to the 20--50\% modulation seen in the light curves (Fig
1).

\begin{figure*}
\begin{center}
\setlength{\unitlength}{1cm}
\begin{picture}(12,11)
\put(16,0){\includegraphics{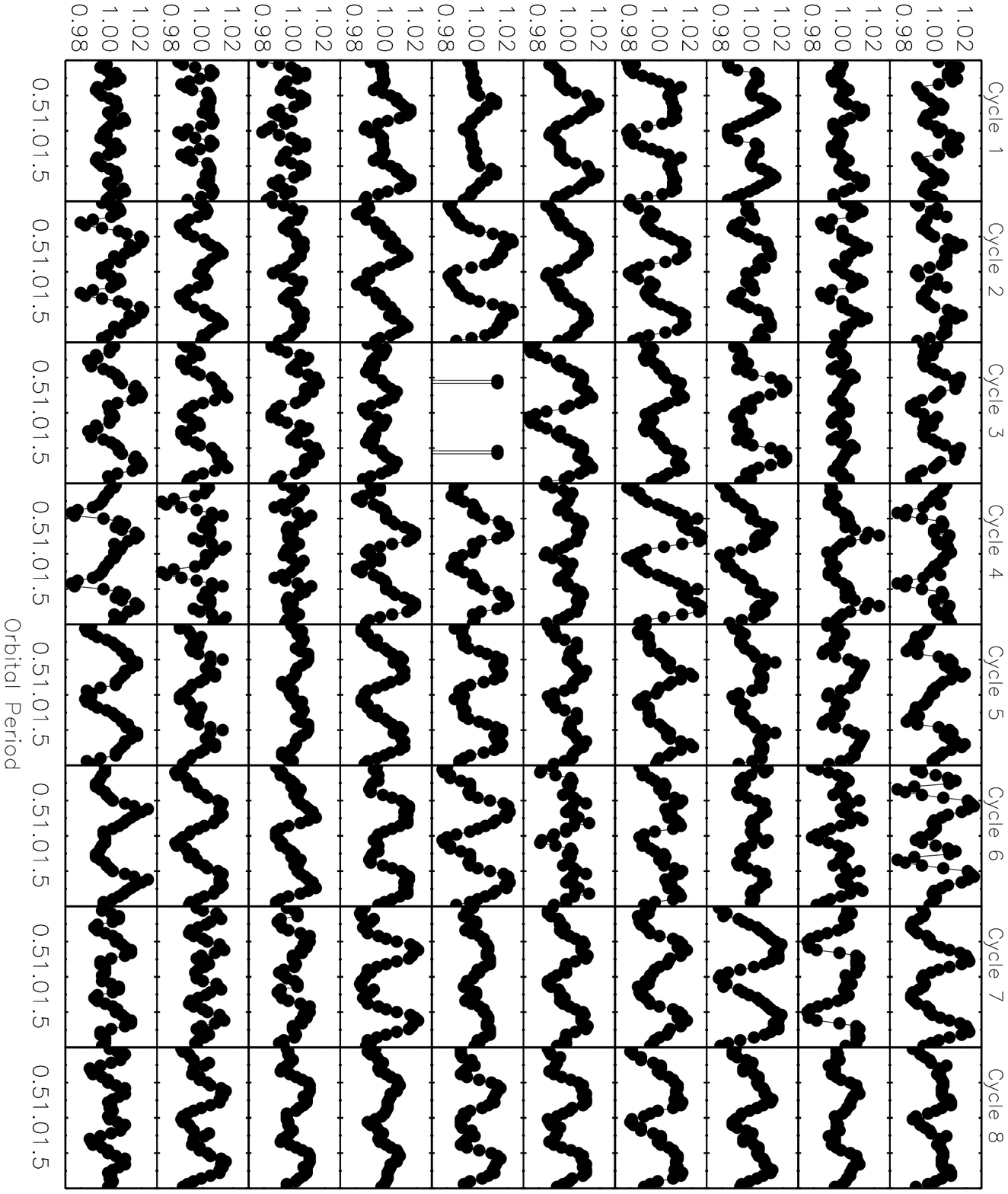}}
\end{picture}
\end{center}
\caption{We have taken the SC data from Q5, removed the trend caused
  by the long period variation, but split up the data so that each
  sub-section contains one long period cycle (arranged from left to
  right). We have then further split these sub-sections into ten bins
  (and arranged from top to bottom) and folded these data on the
  orbital period. (One panel has only two points). It is clear that
  the profile of the folded orbital light curve changes significantly
  over the long cycle.}
\label{scfoldcycle} 
\end{figure*}

Lastly we took the data from Q5 (which has the most complete quarter
of data) and split the data up into long cycles. This was done by
manually selecting the start and end point of each cycle and splitting
this cycle into ten equally phased bins. We then phased the data in
the same way as before and show the resulting folded light curves in
Figure \ref{scfoldcycle}. In the first cycle the light curve
  folded on the orbital period gradually becomes more structured
  having a clear peak in brightness and a peak-to-peak amplitude of
  $\sim$4 percent, before the variation becomes less defined with a
  lower amplitude of variation. Over the next cycles the process
  repeats in a similar, but not identical, manner, with the shape of
  the folded light curve changing over successive orbits.  It is
clear that the data folded on the orbital period can change
significantly in appearance over the long term cycle.

\section{Discussion}

{\src} displays optical photometric variability on two distinct
timescales: the 4 h orbital period and a quasi-period of 10--12 d
which we have termed the long period. As indicated earlier, the 4 h
orbital period is typical of nova-like CVs with bright steady state
accretion disks, which lie predominately above the 2--3 h period
gap. What makes {\src} unusual is the amplitude of the long period
modulation and its phase dependence which does not resemble classical
outbursts seen in CVs.

{\O}stensen et al. (2010) suggested that the long period was due to a
precessing disk. However, the disc would need to have a large tilt, or
alternatively an extended outer rim of varying height, to account for
the large amplitude variation. This would then give rise to negative
super-humps since the accretion disk bright spot should sweep first
across one face of the disk, and then the other and hence varies in
depth in the potential well of the white dwarf (a negative super-hump
has a period a few percent shorter than the orbital period, see Thomas
\& Wood 2015). Positive super-humps show a period longer than the
orbital period by a few percent, but are only seen in systems with a
mass ratio $q=M_{1}/M_{2}$\ltae0.35 (see Wood et al. 2011). For a CV
with a orbital of 4 h, $q$ is not likely to be \ltae0.35 (e.g. Knigge,
Baraffe \& Patterson 2011) and therefore is not expected to show
positive super-humps.

In principal the 4 hr period we observe could be a signature of the
negative super-hump but this period is very steady and the O-C
residuals are very small with no trends. The precession period is a
function of the moment of inertia of the disk, and if the mass
distribution changes, then the precession period will change, and as a
result the negative super-hump period will change, causing variations
in the O-C. These variations are seen in two other CVs observed using
{\kep}, V344 Lyr and V1504 Cyg, where the residuals can show cyclic
changes over the course of the outburst cycle (Osaki \& Kato 2013). We
also note that the amplitude of the super-hump variation in V344 Lyr
is much greater than the amplitude of the orbital variation
(c.f. Still et al. 2010). We therefore consider it highly unlikey that
the long period modulation is due to a precessing disk. 

The profile of the long period modulation (cf Figure \ref{sclight})
shows a relatively slow rise to maximum brightness, whereas dwarf nova
outbursts show a much more rapid increases to maximum (and a greater
amplitude ($>$1 mag).  There are, however, a small group of nova-like
CVs which show `stunted' outbursts with amplitudes up to 1 mag
(e.g. Warner 1995, Honeycutt, Robertson \& Turner, 1995, 1998).
Warner (1995) notes that for a steady state accretion disc to produce
a modulation amplitude of 0.7--1.0 mag would require the mass transfer
rate to change by a factor of ten over the cycle, implying a changing
mass transfer rate was not the cause of the modulation. Honeycutt
(2001) observed nova-like and dwarf nova using a small robotic
telescope and conclude that the cause of the periodic modulation seen
in some nova-like CVs is essentially the same as dwarf nova outbursts.

\begin{figure}
\begin{center}
\setlength{\unitlength}{1cm}
\begin{picture}(6,8)
\put(-1.5,-1.5){\includegraphics{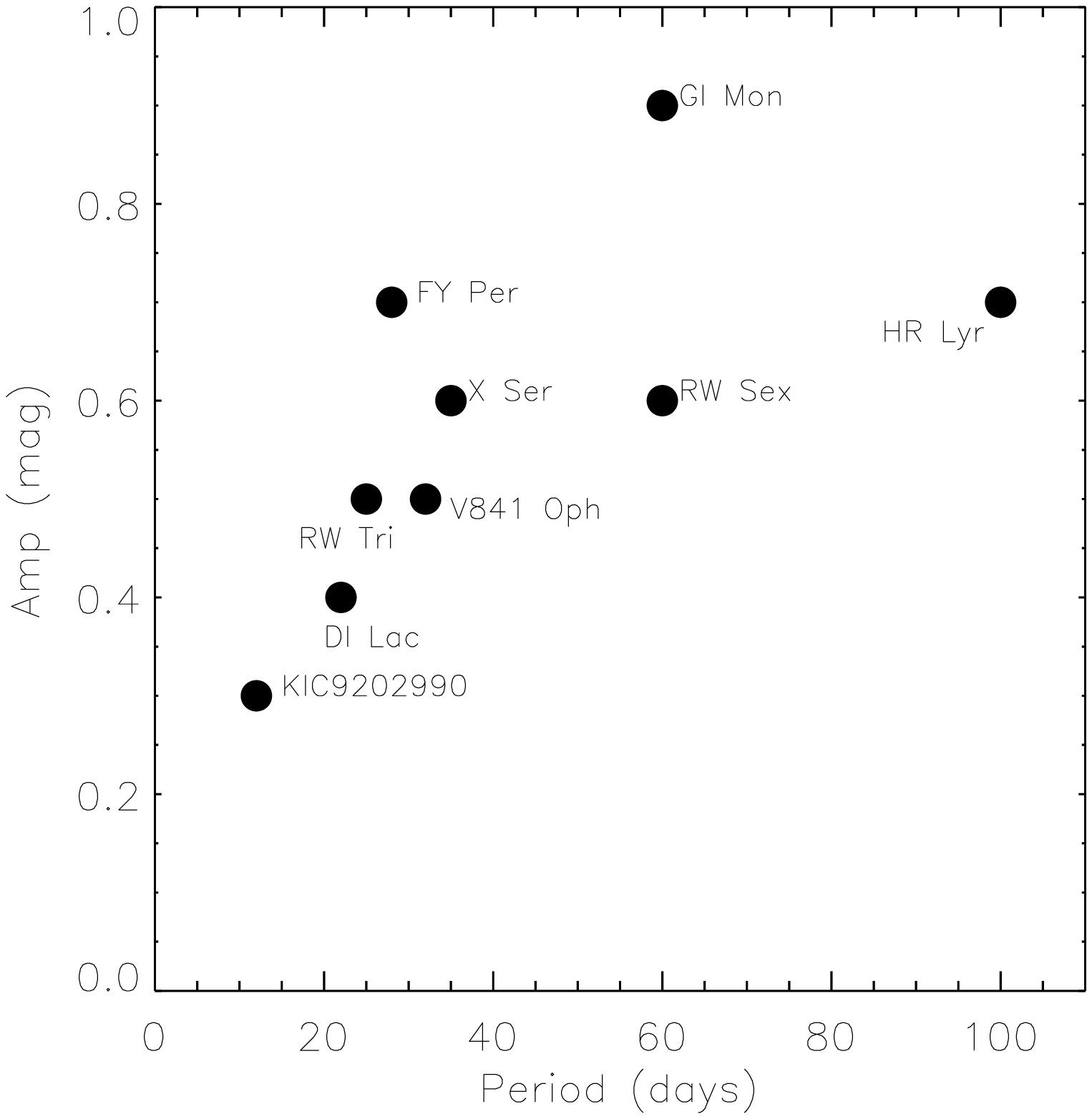}}
\end{picture}
\end{center}
\caption{We compare the amplitude and period of the long period found
  in {\src} with the amplitude and timescale of the repeating stunted
  outbursts of nova-like sources (taken from Honeycutt 2001). For
  comparison KIC 9406652 (Gies et al. 2013) shows $\sim$0.6 mag
  outbursts on a timescale between 27--84 d. {\src} has the shortest
  timescale for stunted outbursts and the smallest amplitude.}
\label{ampperiod} 
\end{figure}

We show in Figure \ref{ampperiod} the relationship between the
amplitude of stunted outbursts and their period of modulation using
the work of Honeycutt (2001) and add the new value for
{\src}. Although there is a clear spread between amplitude and period
of the stunted outburst systems, there is a weak trend with short
periods giving smaller amplitude variations. {\src} is at the short
period end of the distribution and has the lowest amplitude. There is
no correlation between the orbital period and the recurrance time of
the outbursts nor the amplitude of the outburst. {\src} bears some
similarity to a CV also in the {\kep} field, KIC 9406652 (Gies et
al. 2013), which has an orbital period of 6.1 h but also shows
outbursts with amplitude $\sim$0.6 mag on a recurrence timescale of
27--84 d.

Another feature of some of these stunted dwarf novae is the prescence
of `dips' seen in the light curve. Honeycutt, Robertson \& Turner
(1998) show that multiple dips can be seen in systems which have a
depth of 0.2 to $>$0.5 mag with a FWHM ranging from 2 to 50 days. Our
LC observations of {\src} indicate two obvious dips (see Figure 2) of
depths 0.4 and 0.6 mag with FWHM of 2--3 days. These dips share some
resemblance to the dips seen in the nova-like sub-class of CVs the VY
Scl stars, although the duration of these dips are typically tens of
days and are deeper than seen in {\src}. They are thought to be due to
temporary reduction in mass transfer rate which maybe related to
activity on the secondary star (see for instance Howell et al. 2000
and Kafka \& Honeycutt 2005).

\section{Conclusions}

We have explored the photometric properties of {\src} which was
identified as a nova-like CV exhibiting a clear modulation on a 4 h
and a 10--12 d quasi-period using {\kep} data by {\O}stensen et
al. (2010).  We find that the 4 h period is stable and must be the
orbital period. The longer period is not stable and given the
  absence of negative super-humps in the light curve, we consider it
  highly unlikely it is due to the precession of an accretion
  disk. However, we find that the characteristics of this long period
is very similar to the `stunted' outbursts seen in a small number of
nova-like systems. {\src} would be placed on the short period and low
amplitude end of the distribution in these sources.  Medium resolution
optical spectroscopy which could adequately resolve the orbital period
over the course of the long period would allow doppler tomograms to be
made as a function of the long period which would indicate how the
brightness of the disc changes. Another avenue to explore will be to
model the brightness changes over the long term using smoothed
particle hydrodynamical simulation codes such as described for
  instance in Larwood et al. (1996), Simpson (1995), Thomas \& Wood
  (2015).

\section{Acknowledgments}

The data presented in this paper were obtained from the Mikulski
Archive for Space Telescopes (MAST). STScI is operated by the
Association of Universities for Research in Astronomy, Inc., under
NASA contract NAS5-26555. Support for MAST for non-HST data is
provided by the NASA Office of Space Science via grant NNX09AF08G and
by other grants and contracts. This paper includes data collected by
the Kepler mission and is (in part) supported by the National Science
Foundation under Grant No. AST-1305799 to Texas A\&M
University-Commerce and by NASA under grant 11-KEPLER11-0038. Funding
for the Kepler mission is provided by the NASA Science Mission
directorate. This work made use of PyKE (Still \& Barclay 2012), a
software package for the reduction and analysis of Kepler data. This
open source software project is developed and distributed by the NASA
Kepler Guest Observer Office. Armagh Observatory is supported by the
Northern Ireland Government through the Dept of Culture, Arts and
Leisure. We thank the anonymous referee for a helpful report.

\end{document}